\documentclass{revtex4}

\usepackage{graphicx}
\usepackage{amsmath}
\usepackage{dcolumn}
\usepackage{bm}
\usepackage{epstopdf}
\usepackage{multirow} 
\usepackage{array}
\begin{document}
\preprint{APS/123-QED}
 
\title{First multi-reference correlation treatment of bulk metals}

\author{Elena Voloshina$^a$\footnote{E-mail: elena.voloshina@hu-berlin.de}, Beate Paulus$^b$}

\affiliation{
$^a$\mbox{Humboldt-Universit\"at zu Berlin, Institut f\"ur Chemie, 10099 Berlin, Germany}
$^b$\mbox{Freie Universit\"at Berlin, Institut f\"ur Chemie und Biochemie, 14195 Berlin, Germany}
}

\begin{abstract}
Existence of the $sp$-$d$ hybridization of the valence band states of the \textit{fcc} Ca and Sr in the vicinity of the Fermi level indicates that their electronic wave function  can have a multi-reference (MR) character. We performed a wave function-based correlation treatment for these materials by means of the method of increments. As oppose to the single-reference correlation treatment (here: coupled cluster), which fails to describe cohesive properties in both cases, employing the MR averaged coupled pair functional one can achieve almost $100$\,\% of the experimental correlation energy. 
\end{abstract}

\maketitle

\section{Introduction}

\textit{Ab initio} Hartree-Fock (HF) and post-HF electron correlation methods, such as M\o ller-Plesset perturbation theory (MP) and  coupled cluster theory (CC), are standard tools in computational chemistry nowadays and various program packages are available for accurate calculations of properties of atoms and molecules. For solids, HF calculations are possible with, e.g., the program package CRYSTAL~\cite{crystal}. However, the problem of an accurate treatment of electron correlation is not fully settled. 

Various approaches have been devised that attack the problem directly. When focusing on MP2, examples include the periodic atomic-orbital-based MP2 method by Scuseria and co-workers~\cite{scuseria}, the local and density-fitted MP2 approach implemented in CRYSCOR~\cite{cryscor}, and the plane-wave-based MP2 code developed in VASP~\cite{kresse}. In fact, in VASP, methods beyond MP2 are being developed and applied to a variety of real solids~\cite{fciqmc}. However all these applications are dealing with insulating systems.

An alternative way is to treat electron correlation through considering finite clusters. Here two variations are possible. Firstly, one may gradually approach the infinite system starting from finite clusters of increasing size, which are tractable using standard quantum-chemical correlation methods. For the transfer of information from the finite clusters to the infinite solid, an extrapolation scheme can be employed. Such schemes may be based on the spatial extensions of the clusters~\cite{cluster1}.  Unfortunately, the cluster size needed for a reliable extrapolation is rather large, and this may be prohibitive for the application of post-HF methods. Other schemes achieve a more rapid convergence by combining high-level (e.g., MP2) results for clusters with lower-level (e.g., DFT) results for clusters and solid~\cite{cluster2,cluster3}.  

A branch of approaches is inspired by the many-body expansion (see e.g. Refs.~\cite{mb1,mb2}). One of the approaches of this type is the \textit{method of increments}, originally proposed by Stoll~\cite{stoll, stoll2} and further developed by other groups~\cite{paulus,ce,h2o,fried,fried2,mata,carsten1}. In this approach, a periodic HF calculation is followed by a many-body expansion of the correlation energy, where the individual units of the expansion are either atoms or other domains of localized orbitals~\cite{carsten2}. Calculations based upon the method of increments have been performed on a variety of solids with band gaps (for a review, see Ref.~\cite{paulus}).  In the previous publications~\cite{mg1,mg2,zn,hg,review} we have shown that the incremental scheme, after some reformulation,  can also be applied when considering metals. In the systems studied so far the degeneracy at the Fermi level was not so high and a single-reference (SR) quantum-chemical treatment (here: CC) was therefore still sufficient. The questions arises how one should deal with systems where a SR treatment breaks down, i.e.\,\,the strongly correlated cases with high degeneracy at the Fermi level.

The first publication considering such a situation is related to the correlation energy of a metallic Li-rings~\cite{li-rings}. Later an embedding scheme was suggested for three-dimensional lithium relying on pairs of atoms~\cite{li-embed}. However, up to now there was no systematic application of the multi-reference (MR) incremental scheme to evaluate cohesive properties of a bulk metal. Here we perform such a study taking calcium and strontium as examples. 

Under normal conditions calcium and strontium crystallize into highly symmetric close packed structure with one atom per unit cell, a face-centered cubic (\textit{fcc}) structure [Figure \ref{fig:ca-fcc} (a)]. These metals are electronically fairly simple: below the Fermi level ($E_F$) the electronic bands are practically free-electron-like, while the broad and almost unoccupied $d$ bands are slightly above $E_F$. As a consequence, near $E_F$ the electronic wave functions are modified by hybridization with the $d$ bands. The aim of the present study is to check whether single-reference treatment is still sufficient as degree of $sp$-$d$ hybridization is rather low~\cite{ca-hf} and to investigate how MR incremental scheme will behave in such a case.

\section{\label{comput}Theoretical and computational details}

\subsection{Method of increments for metals}

The method of increments combines HF calculations for periodic systems with correlation calculations for finite embedded fragments, reflecting the lattice structure of the system under study, and the total correlation energy per unit cell of a crystal is written as a cumulant expansion in terms of contributions from localized orbital groups of increasing size:
\begin{equation}\label{eqn:corr}
E^\mathrm{corr.}_\mathrm{crystal} = \sum_{i\,\,\in\,\,u.c.}\varepsilon_{i}+\frac{1}{2!}\sum_{\begin{subarray}{l}i\,\,\ne\,\,j\\ i\,\,\in\,\,u.c.\\ j\,\in\,\mathrm{crystal}\end{subarray}}\Delta\varepsilon_{ij}+ \frac{1}{3!}\sum_{\begin{subarray}{l}i\ne j\ne k\\ i\,\,\in\,\,u.c.\\ j,k\,\in\,\mathrm{crystal}\end{subarray}}\Delta\varepsilon_{ijk}+ ...\quad .
\end{equation}
Here, the summation over $i$ involves orbital groups located in the reference cell, while those over $j$ and $k$ include localized orbital groups from all the centers of the crystal. The $\varepsilon_i$ (one-body increment) is computed by considering excitations only from the $i$-orbitals, freezing the rest of the solid at the HF level. The non-additive two-body contribution is defined as $\Delta\varepsilon_{ij} = \varepsilon_{ij}-(\varepsilon_i +\varepsilon_j)$ where $\varepsilon_{ij}$ is the correlation energy of the joint orbital system $(ij)$. Higher-order increments are defined in an analogous way. A detailed description of this approach can be found in Ref.~\cite{paulus} In this section we outline briefly how this method can be applied for the calculation of ground-state properties of metals.

The presence of well-localized orbitals is a precondition for the applicability of local correlation schemes. For metals, however, it is no longer possible to generate well-localized Wannier orbitals by unitary transformation within the occupied HF space. To overcome this difficulty it has been suggested to base the incremental expansion on a well-localizable model system rather than on the real metal and to allow for delocalization (and thereby approach the real metal) only gradually within the various levels of the many-body expansion~\cite{embed}. The related procedure, allowing to localize the orbitals in metals, consists of several steps. Firstly, one performs preliminary HF calculation with a minimal valence basis set for the whole cluster consisting from the atoms to be correlated (in the centre of the cluster) and the embedding atoms (see Sec.~II\,C). This way, delocalization is avoided and no metallic character can be described. However, each atom still has its correct crystal surroundings with respect to the electrostatic interaction. Due to the neglect of any metallic character one can use standard procedures to localize the orbitals. The resulting set of localized orbitals contains both the embedding orbitals which are centered at the embedding atoms and also the orbitals located at the atoms which are to be correlated. Therefore, in the next step, one separates between the orbitals of the central region and the embedded ones and supplies on the atoms to be correlated an extended basis set. After recalculating the integrals and performing SCF calculations for the central part of the cluster, one can use the reoptimized orbital set in  correlation calculations with any size-extensive method. For further details, see Ref.~\cite{embed}.

\subsection{Multi-reference incremental scheme}

\begin{figure}[h]
\includegraphics[width=0.96\textwidth]{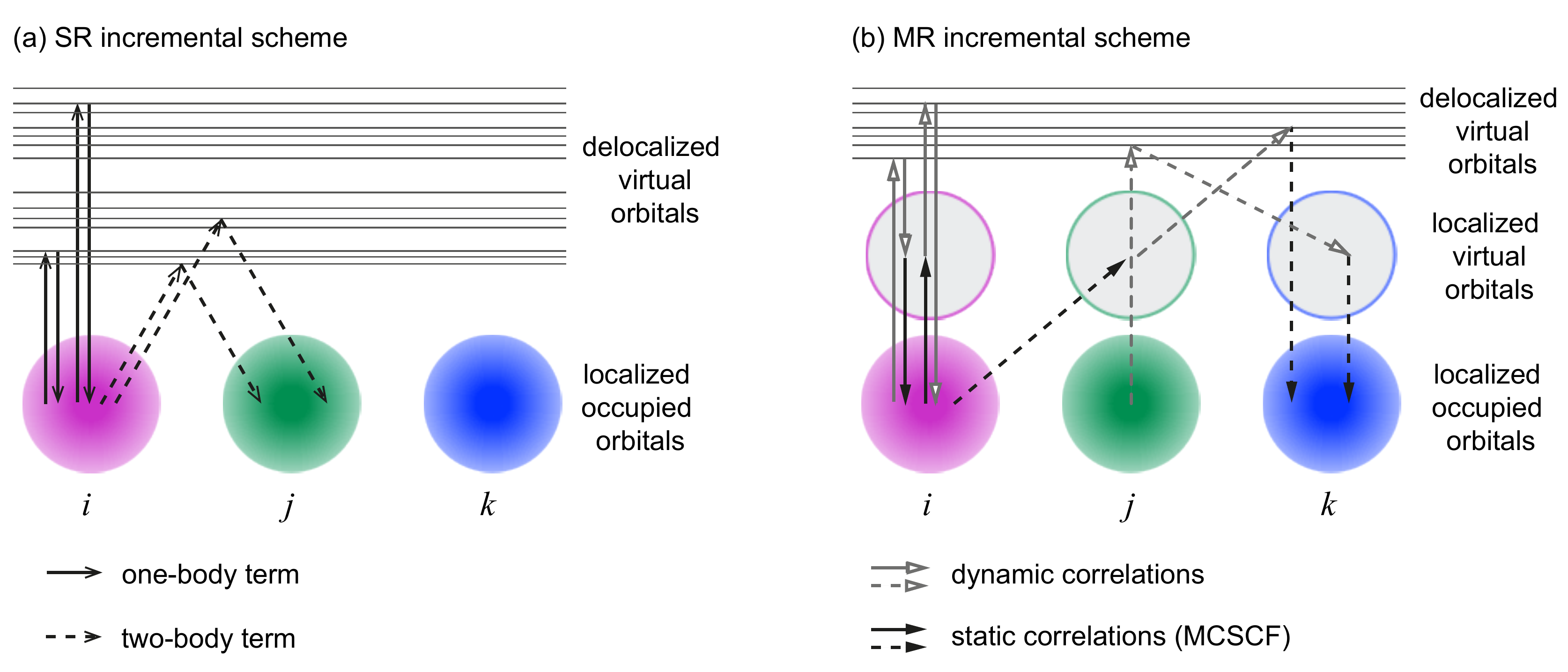}
\vspace{0.05cm}
\caption{Schematic view of the single-reference (a) and multi-reference (b) incremental scheme. }
\label{fig:scheme}
\end{figure}

Within the MR approach, the correlation energy is partitioned into a complete active space self-consistent field (CASSCF) part, which describes the static correlations, and a MR configuration interaction part for the dynamic correlations. Contrary to the SR approach, where only occupied orbitals are localised [Figure \ref{fig:scheme} (a)], also the part of the virtual orbitals, which are treated in the MCSCF procedure, is localised [Figure \ref{fig:scheme} (b)]~\cite{paulus}. An active space for the CASSCF calculation includes the important bands around the Fermi level. The higher-lying virtual orbitals are kept delocalised. A compromise has to be found between accuracy (large active space) and computational effort. In the case of Ca and Sr, our tests show, that it is enough to use $4$ active orbitals per atom, rather than $9$ as may be expected: If we increase the active space over $4$ active orbitals per atom, the occupation numbers of the additional orbitals are less than $0.05$ electrons and therefore negligible in the CASSCF treatment. 

The incremental expansion looks formally identical to the single-reference case, but the correlation treatment to determine, e.g., the one-body increment, $\varepsilon_i$, is different. The reference energy is the single-reference HF energy of the system. The static correlations within the occupied and unoccupied orbitals of centre $i$ are calculated on a complete-active-space CI level where the active space is built up from the orbitals at centre $i$. The dynamic correlations are to be treated on top of the CASSCF wave function with a MR correlation method. Excitations are possible into the whole delocalised virtual space. Step by step more orbital groups are correlated simultaneously and the increments are calculated according to Eq.~\ref{eqn:corr}. In the case of metallic systems, one is dealing with two types of contributions to the electronic correlations: The short-range part, which can be treated with the multi-reference version of the method of increments, and the long-range fluctuations, which are determined by a charge screening and can be approximated by a random-phase approximation~\cite{rpa-incr}. The latter was not performed in the present study.

An important issue is the choice of the quantum-chemical method to be used for evaluating the increments. In the incremental scheme the total correlation energy is divided into individual contributions and it is therefore important that the correlation method applied is size-extensive. For the SR case a variety of size-extensive methods is available, especially the CCSD(T) method~\cite{ccsd} is widely used and yields very good results for the correlation energy. For the MR case the problem exists, that the available multi-reference correlation methods are not strictly size-extensive. The MCSCF part of the calculations, which corresponds to a full CI in a restricted space with optimization of the orbitals, is still size-extensive. The problem arises for the correlation treatment performed on top of it. In order to achieve an approximate size-extensivity, there are two possibilities: One is a correction of the correlation energy obtained with MRCISD after the CI calculation. The two most common corrections are by Davidson~\cite{dav} and by Pople~\cite{pop}. A more elaborate correction is proposed by Gdanitz and Ahlrichs~\cite{gdahl}. There the correction takes place within the CI iterations themselves. In our studies we have use the MR averaged coupled pair functional (MRACPF)~\cite{gdahl}, an approximately size-extensive MR correlation method, as it is implemented in the program package MOLPRO~\cite{mracpf,molpro}, where it can be used with different separation of internal and external space. According to our tests performed for Ca and Sr, correlation-energy increments obtained by means of MRACPF perfectly agree with the results of MRCI with Pople correction. 

All correlation calculations are performed with the program package MOLPRO\cite{molpro}. 

\subsection{Structural models and basis sets}

\begin{figure}[h]
\vspace{0.2cm}
\includegraphics[width=0.96\textwidth]{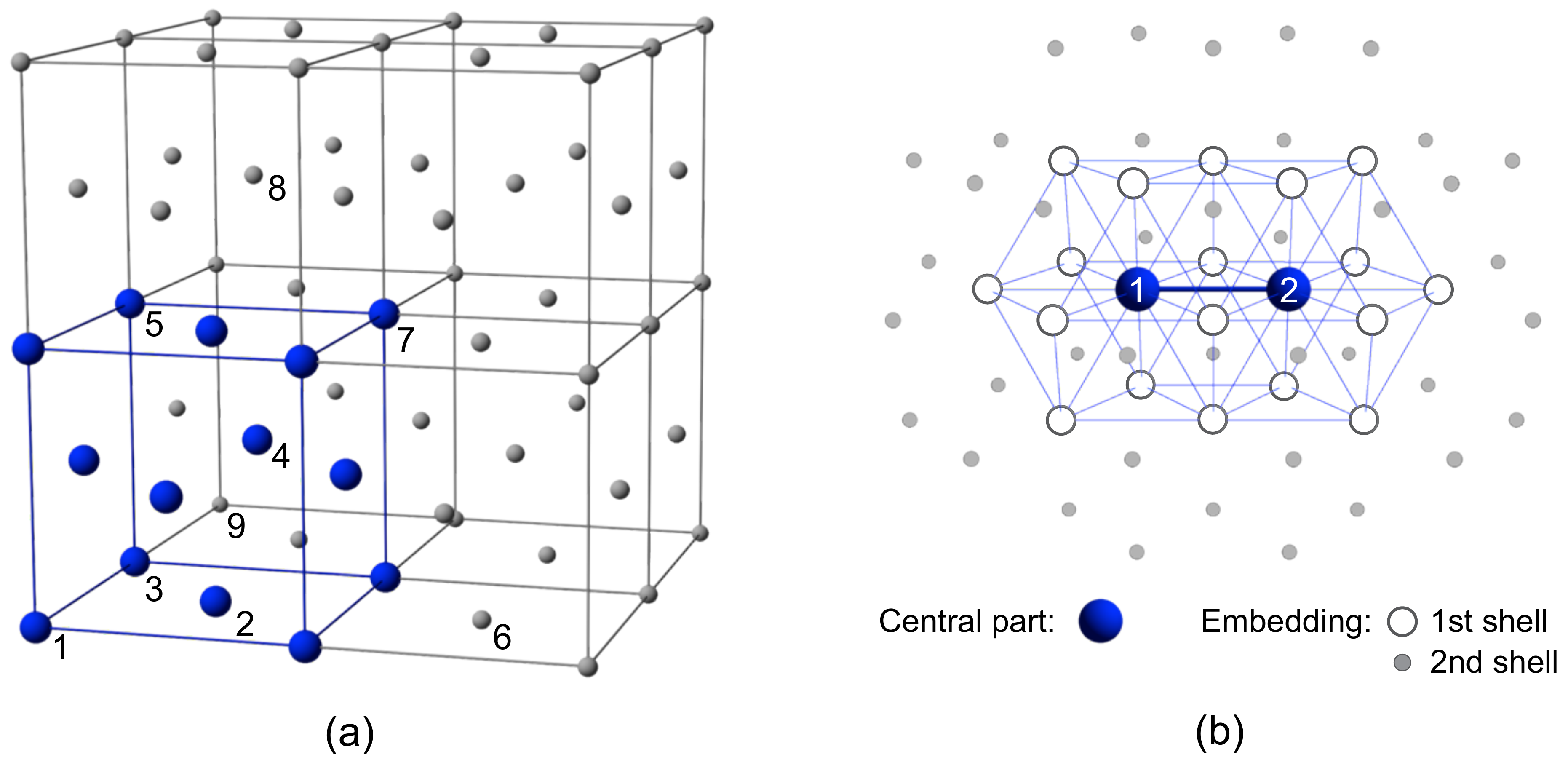}
\vspace{0.2cm}
\caption{(a) Crystallographic structure of the \textit{fcc} Ca and Sr. Large dark spheres show the unit cell. The atoms are ordered by their distance from the atom number 1. (b) One of the so-called two-body clusters, a cluster for calculating the two-body contribution to the correlation energy ($r_{12}=a/\sqrt{2}$). Spheres of different sizes and styles represent atoms of different types (see text for details). }
\label{fig:ca-fcc}
\end{figure}

The correlation energy increments are calculated for selected fragments which reflect the lattice structure of the studied system [Figure \ref{fig:ca-fcc} (a)]. These fragments have two components: first the atoms $i$, $j$, $k$, ..., to be correlated (in the center of the fragment), and second the embedding atoms. We select the positions of the atoms to be  correlated and surround each of these atoms by an embedding shell with a cutoff for the embedding of two times the lattice parameter, $a$, of the system under study. This leads to the cluster consisting of $43$ atoms ($1$ atom in the central part and $42$ embedding atoms), when considering the one-body contribution. To get the results presented in this work the embedding was partitioned into two regions: a true embedding region (the second embedding shell), where the atoms are described with a minimal basis, and an intermediate region (the first embedding shell), where the occupied orbitals are kept frozen but additional basis functions are supplied for the reoptimization of the orbitals in the central part of the cluster.\cite{be,mg2} An example of the embedded two-body cluster is shown in Figure \ref{fig:ca-fcc} (b).

The Ca and Sr atoms of the central region are considered with small-core pseudotentials ($\mathrm{Ca^{10+}}$-PP and $\mathrm{Sr^{10+}}$-PP)\cite{pp1} and the related basis sets augmented by $f$ polarization functions ($\alpha_f =0.151$\,$\mathrm{Bohr}^{-2}$  and $0.135$\,$\mathrm{Bohr}^{-2}$, optimized for Ca and Sr atoms, respectively) up to $(6s6p5d1f)/[4s4p2d1f]$. Embedding atoms are described with the two-valence electron pseudopotentials~\cite{pp2} and either the corresponding basis set of valence double-$\zeta$ quality\cite{pp2} [the first embedding shell; see Figure \ref{fig:ca-fcc} (b)] 
or a minimal (4s)/[1s] part of the basis set from Ref.~\cite{pp2} [the second embedding shell; see Figure \ref{fig:ca-fcc} (b)]. 

\section{Results and discussion}

A starting point for the treatment of the many-body correlation effects in solids is a reliable HF self-consistent-field result for the infinite system. Such data for Ca and Sr, i.e. the HF cohesive energies ($E_\mathrm{coh}^\mathrm{HF}$), can be found in the literature~\cite{ca-hf} and are $-0.58$\,eV/atom and $-0.44$\,eV/atom, respectively. These data significantly underestimated relatively to the experimental values of $-1.84$\,eV/atom~\cite{kittel} and $-1.72$\,eV/atom\cite{kittel} known for Ca and Sr, respectively.

The cohesive contribution of the one-body increment is defined as the difference between the correlation energy of the embedded atom ($\varepsilon_i$) and the free atom ($E^\mathrm{corr}_\mathrm{free}$): $\Delta\varepsilon_{i}=\varepsilon_{i}-E^\mathrm{corr}_\mathrm{free}$. The basis set superposition error was corrected by applying the counterpoise method,\cite{bsse} considering $42$ ghost atoms. For both systems under study, the one-body contributions are small and repulsive: $\Delta\varepsilon_i\mathrm{(Ca)}=87$\,meV and $\Delta\varepsilon_i\mathrm{(Sr)}=24$\,meV (see also Tables \ref{ca-corr} and \ref{sr-corr}).

The incremental expansion makes sense only if the number of non-negligible terms is sufficiently small, i.e. if the number of centers to be treated simultaneously for determining non-additive corrections is, say, three or four at most, and if the magnitude of all these corrections decreases rapidly with increasing spatial distance between the centers. Before checking whether the incremental expansion fulfils these convergence criteria, one has to define the increments.

\begin{table}[t]
\caption{Correlation-energy contributions to the cohesive energy of the \textit{fcc} Ca obtained as discussed in the text, in eV per atom.}
\label{ca-corr}
\begin{tabular}{p{3.5cm}cp{2.8cm}p{2.8cm}p{2.8cm}p{1.8cm}}
\hline
\quad\quad\,\,\,\,Method:					&&CCSD  &CCSD(T)  & MCSCF& MRACPF   \\
\hline
\quad\quad\,\,\,\,$\Delta\varepsilon_i$			&&$+0.087$&$+0.087$&$+0.091$&$+0.087$\\
&&&&&\\
Total $\sum\,\Delta\varepsilon_{ij}$			&&$-1.951$&$-2.587$&$-0.223$&$-2.855$\\
&&&&&\\
\quad\quad\,\,\,\,$\sum\Delta\varepsilon_{ijk}^{a}$		&&$+1.196$&$+0.874$&$-1.294$&$+1.713$\\
\quad\quad\,\,\,\,$\sum\Delta\varepsilon_{ijk}^{b}$		&&$ -0.326$&$ -0.534$&$ -0.418$&$-0.030$\\							
Total $\sum\,\Delta\varepsilon_{ijk}$			&&$+0.870$&$+0.339$&$-1.712$&$+1.682$\\
&&&&&\\
\quad\quad\,\,\,\,$\sum\Delta\varepsilon_{ijkl}^{[4]}$		&&$ -1.185$&$ -1.192$&$-0.613$&$-0.382$\\
\quad\quad\,\,\,\,$\sum\Delta\varepsilon_{ijkl}^{[3]+1}$	&&$+0.386$&$+0.406$&$+0.196$&$+0.153$\\
\quad\quad\,\,\,\,$\sum\Delta\varepsilon_{ijkl}^{[2]+[2]}$	&&$+0.134$&$+0.129$&$+0.088$&$+0.084$\\
Total $\sum\,\Delta\varepsilon_{ijkl}$			&&$ -0.665$&$ -0.657$&$-0.332$&$-0.145$\\
&&&&&\\
\quad\quad\,\,\,\,$E_\mathrm{coh}^\mathrm{corr}$		&&$ -1.659$&$-2.818$&$-2.176$&$-1.231$\\
\hline
\end{tabular}

\vspace{0.2cm}
\hspace{-5.5cm}
\footnotesize{$^a$Increments belonging to the groups $A$, $B$, and $C$ (see text for details)}; 

\hspace{-10.35cm}
\footnotesize{$^b$The rest of the three-body increments}.
\end{table}

\begin{table}[t]
\caption{Correlation-energy contributions to the cohesive energy of the \textit{fcc} Sr obtained as discussed in the text, in eV per atom.}
\label{sr-corr}
\begin{tabular}{p{3.5cm}cp{2.8cm}p{2.8cm}p{2.8cm}p{1.8cm}}
\hline
\quad\quad\,\,\,\,Method:					&&CCSD  &CCSD(T)  & MCSCF& MRACPF   \\
\hline
\quad\quad\,\,\,\,$\Delta\varepsilon_i$			&&$+0.024$&$+0.024$&$+0.018$&$+0.024$\\
&&&&&\\
Total $\sum\,\Delta\varepsilon_{ij}$			&&$-1.839$&$-2.560$&$-0.222$&$-2.896$\\
&&&&&\\
\quad\quad\,\,\,\,$\sum\Delta\varepsilon_{ijk}^{a}$		&&$+1.053$&$+0.709$&$-1.486$&$+1.695$\\
\quad\quad\,\,\,\,$\sum\Delta\varepsilon_{ijk}^{b}$		&&$ -0.253$&$ -0.423$&$-0.313$&$+0.027$\\
Total $\sum\,\Delta\varepsilon_{ijk}$			&&$+0.799$&$+0.286$&$-1.799$&$+1.772$\\
&&&&&\\
\quad\quad\,\,\,\,$\sum\Delta\varepsilon_{ijkl}^{[4]}$		&&$ -1.016$&$ -0.998$&$ -0.502$&$-0.322$\\
\quad\quad\,\,\,\,$\sum\Delta\varepsilon_{ijkl}^{[3]+1}$	&&$+0.231$&$+0.247$&$+0.144$&$+0.112$\\
\quad\quad\,\,\,\,$\sum\Delta\varepsilon_{ijkl}^{[2]+[2]}$	&&$+0.124$&$+0.098$&$+0.078$&$+0.074$\\
Total $\sum\,\Delta\varepsilon_{ijkl}$			&&$ -0.661$&$ -0.653$&$ -0.280$&$-0.136$\\
&&&&&\\
\quad\quad\,\,\,\,$E_\mathrm{coh}^\mathrm{corr}$		&&$ -1.677$&$-2.903$&$-2.283$&$-1.236$\\
\hline
\end{tabular}

\vspace{0.2cm}
\hspace{-5.5cm}
\footnotesize{$^a$Increments belonging to the groups $A$, $B$, and $C$ (see text for details)}; 

\hspace{-10.35cm}
\footnotesize{$^b$The rest of the three-body increments}.

\end{table}

\begin{figure}[h]
\includegraphics[width=0.96\textwidth]{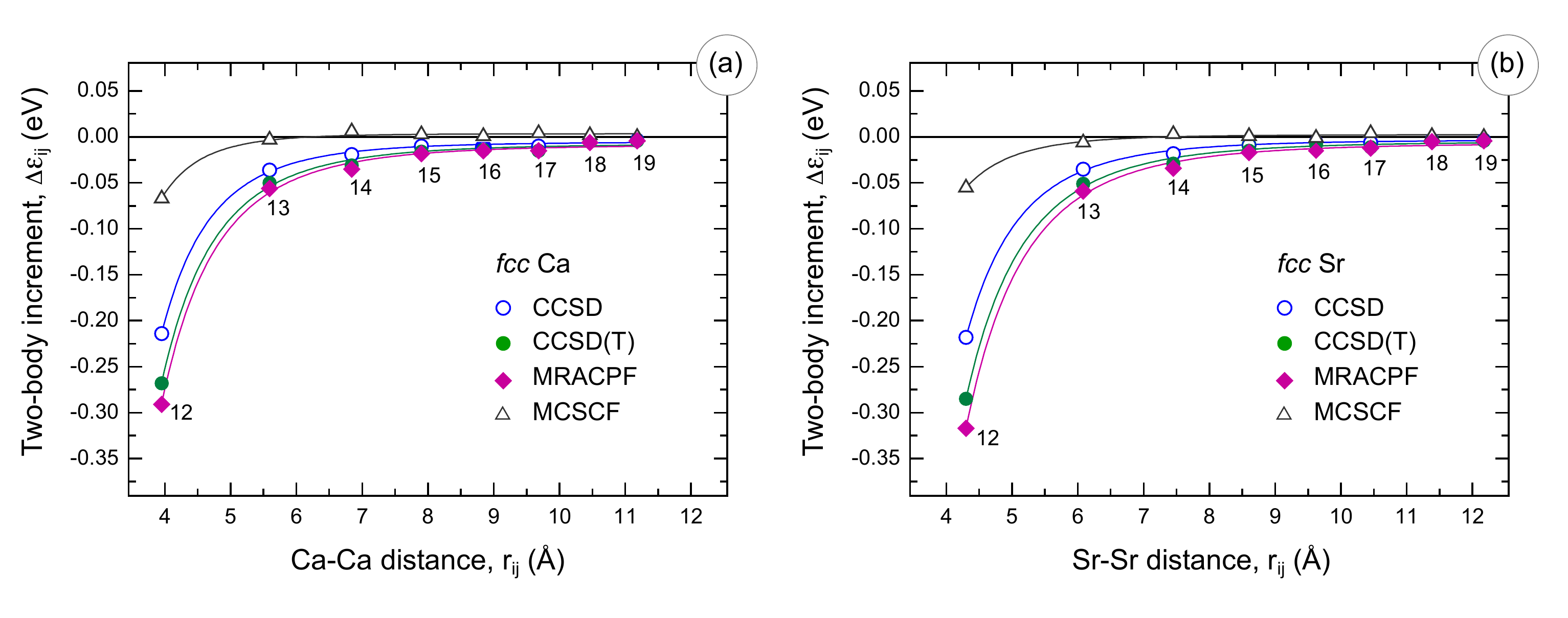}
\caption{The two-body correlation energy  increments are plotted as a function of distance, for both calcium (a) and strontium (b). The solid lines are eye-guides. }
\label{fig:2body}
\end{figure}

The two-body increments may in principle be simply defined by a distance cutoff as given in Table \ref{tbl:2body}, where the two-body increments are ordered by distance within the \textit{fcc} lattice [Figure \ref{fig:ca-fcc} (a)]. So long as the energy is dominated by a rapid decay with distance then the use of a simple cutoff is well justified. As can be seen from Figure \ref{fig:2body},  the latter condition is met for both Ca and Sr independently on the quantum-chemical approach employed, although the weighted sum of individual increments varies from $-1.951$\,eV (for CCSD) to $-2.855$\,eV (for MR-ACPF) in the case of Ca (see Table \ref{ca-corr}) and from $-1.839$\,eV (for CCSD) to $-2.896$\,eV (for MR-ACPF) in the case of Sr (see Table \ref{sr-corr}). 

\begin{table}[t]
\caption{Definition of the two-body increments for Ca and Sr. The two-body increments are ordered by distance within the \textit{fcc} lattice as given in Figure \ref{fig:ca-fcc} (a). The distances are given both in units of $a$, the lattice parameter, and in \AA\ (at the experimental lattice parameters). Weight factors show how often the individual increments occur in the considered structure.}
\label{tbl:2body}
\begin{tabular}{p{3cm}p{3cm}p{3cm}p{3cm}cp{1.5cm}}
\hline
Label & $r_{ij}$ ($a$)		&\multicolumn{2}{c}{$r_{ij}$ (\AA)\hspace{1cm} }	&&Weight\\
\cline{3-4}
($ij$)  &				&\textit{fcc} Ca	&\textit{fcc} Sr 	&& factor\\         
\hline
12	&$1/\sqrt{2}$          	&$ 3.95$	&$ 4.30$	&&\hspace{0.35cm}$6$\\
13	&$1$                        	&$ 5.59$	&$ 6.08$	&&\hspace{0.35cm}$3$\\
14	&$\sqrt{3/2}$		&$ 6.84$	&$ 7.45$	&&\hspace{0.35cm}$12$\\
15	&$\sqrt{2}$          	&$ 7.90$	&$ 8.60$	&&\hspace{0.35cm}$6$\\
16	&$\sqrt{5/2}$		&$ 8.84$	&$ 9.62$	&&\hspace{0.35cm}$12$\\
17	&$\sqrt{3}$     		&$ 9.68$	&$10.54$	&&\hspace{0.35cm}$4$\\
18	&$\sqrt{7/2}$		&$10.46$	&$11.38$	&&\hspace{0.35cm}$24$\\
19	&$2$			&$11.18$	&$12.17$	&&\hspace{0.35cm}$3$\\
\hline
\end{tabular}
\end{table}

\begin{figure}[t]
\includegraphics[width=0.92\textwidth]{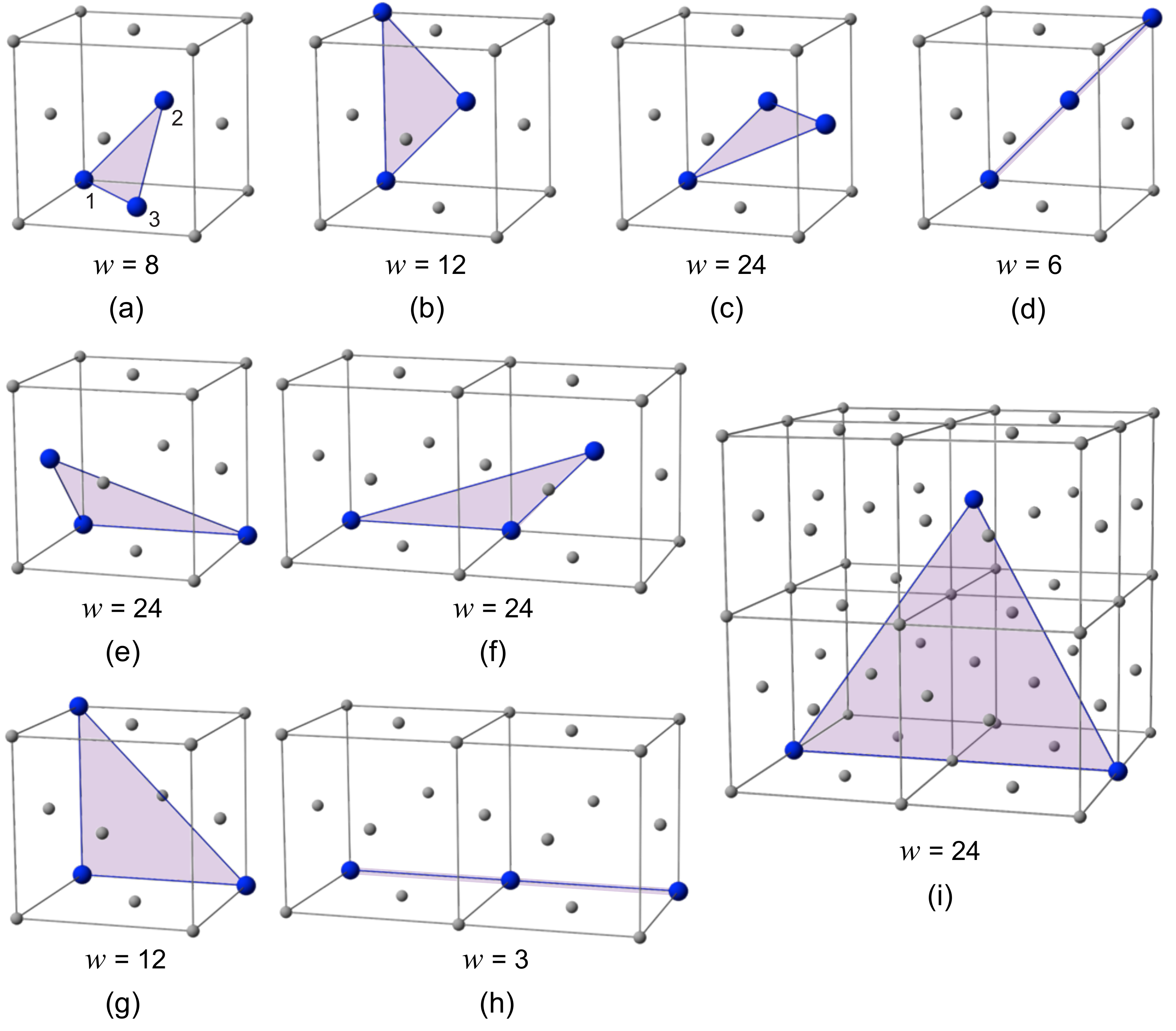}
\caption{Central parts of the selected three-body clusters (large spheres) are shown on a background of an \textit{fcc} cell. The weight factors of the increments ($w$) in the \textit{fcc} structure are given.}
\label{fig:3body1}
\end{figure}

For the higher-order increments the choice of cutoff criteria is more complicated. When adopting the same criteria as in the case of the two-body increments, one gets $48$ different three-body clusters with metal-metal distances in the central part, which are not longer than $2a$, where $a$ is the lattice constant of the \textit{fcc} Ca or Sr. Central parts of the most compact and most expanded three-body clusters are shown in Figure \ref{fig:3body1} (a) and (i), respectively.  As expected, all individual three-body increments are smaller in magnitude than the corresponding two-body increments in the incremental expansion. Even the sum of all three-body increments is smaller than the sum of the two-body. It is important, that difference between the individual correlation-energy increments computed with CC methods and MRACPF is not higher than $10$\,meV (for the most compact arrangements). However due to the large weight factors the total three-body sum varies significantly with the method employed (see Tables \ref{ca-corr} and \ref{sr-corr}). 

\begin{figure}[t]
\includegraphics[width=1.00\textwidth]{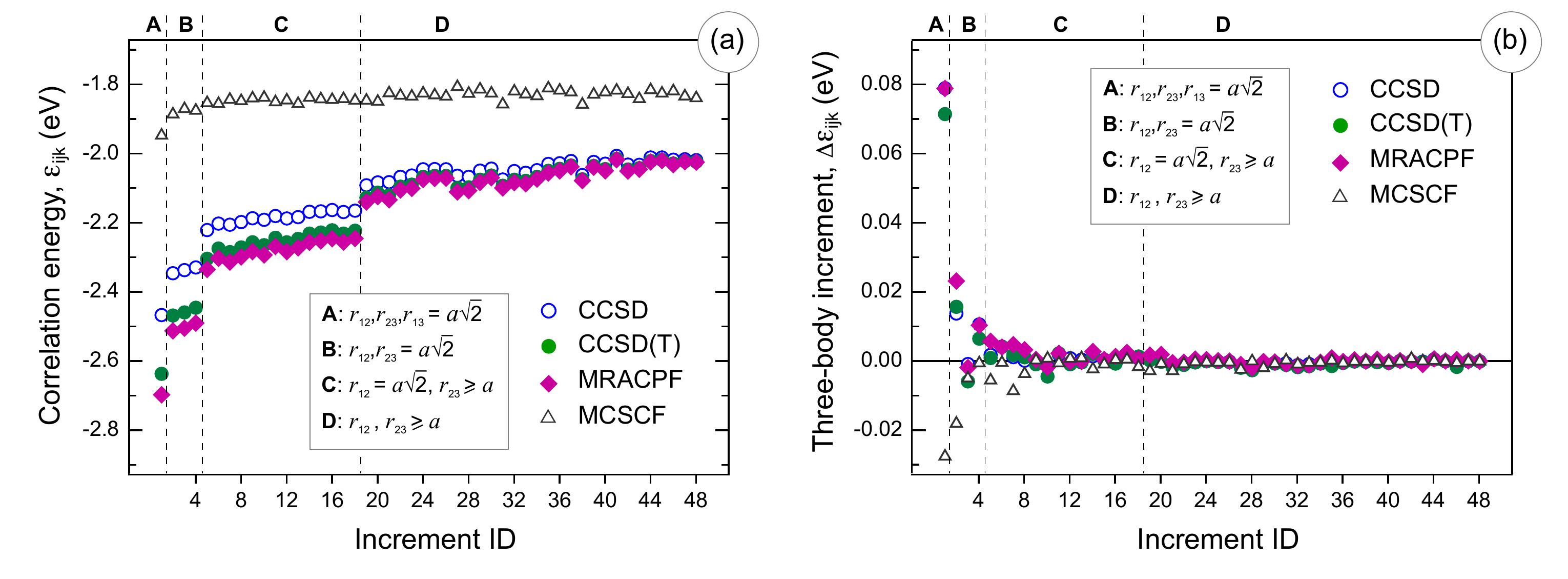}
\caption{Correlation energy contributions ($\varepsilon_{ijk}$ and $\Delta\varepsilon_{ijk}$) computed for Ca are plotted for the $48$ considered three-body species. The increments are ordered according to increasing amount of metal-metal distances larger than $a/\sqrt{2}$ in the central part of studied cluster: This way four groups, \textit{A}-\textit{D}, can be evaluated  (indicated with dashed lines). If the amount of nearest neighbours is equal for several species within the same group, their correlation energies are plotted as functions of an arithmetic mean of the distances between metal atoms of the central part, $\bar{r}=\sum r_{ij} /3$. The pictures obtained for Sr are qualitatively the same.}
\label{fig:3body2}
\end{figure}

The correlation energy, $\varepsilon_{ijk}$, depends strongly on the \textit{compactness} of the three-body cluster. As follows from Figure \ref{fig:3body2} (a), one can easily subdivide all considered three-body species into four groups, $A$-$D$, according to the growing amount of metal-metal distances in the central part larger than $r_{NN}=a/\sqrt{2}$. The weight of static correlations increases when moving from  \textit{A} to \textit{D}, taking values between $70$ and $90$\,\%. The behaviour of increments, $\Delta\varepsilon_{ijk}$, is more complicated and harder to systematise. Obviously, individual values depend strongly on the geometry of the central part and while \textit{acute triangles} tend to yield positive energies, $\Delta\varepsilon_{ijk}$ obtained for \textit{obtuse triangles} are attractive. The same conclusion was made when studying the other group 2 and 12 metals.\cite{mg1,zn} One can see, when analysing Figure \ref{fig:3body2} (b), that the more compact arrangements (groups \textit{A}-\textit{C}) lead to the contributions which are higher in magnitude than the values obtained for the expanded structures (group \textit{D}). 
Neglecting the increments of group \textit{D}, yields deviation in $\sum\Delta\varepsilon_{ijk}$ obtained with the MRACPF which is lower than $2$\,\% (compare ``$\sum\Delta\varepsilon_{ijk}^{a}$'' and ``Total $\sum\Delta\varepsilon_{ijk}$'' in Tables~\ref{ca-corr} and \ref{sr-corr}, for Ca and Sr, respectively). This means that it is not necessary to use the same cutoff criteria for the second and third order increments and only those of the groups A, B, and C have to be taken into account in the latter case. [Note: The situation is different at the CC level. Here the members of the group $D$ contribute with at least $30$\,\% (for the CCSD) to the total sum of three-body increments]. 

\begin{figure}
\includegraphics[width=0.92\textwidth]{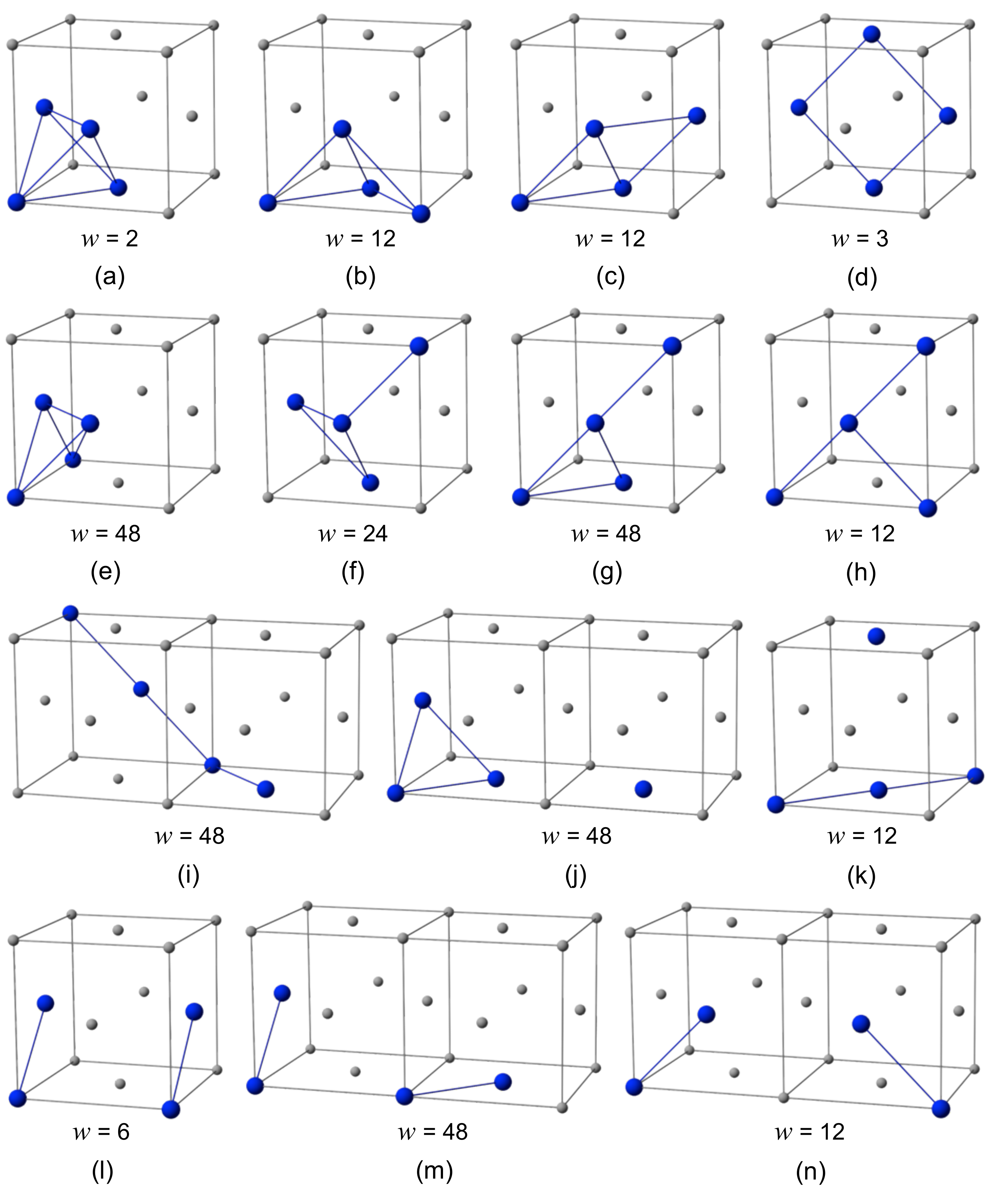}
\caption{Central parts of the selected four-body clusters (large spheres) are shown on a background of an \textit{fcc} cell with all bonds of length $a/\sqrt{2}$ drawn. The weight factors of the increments ($w$) in the \textit{fcc} structure are given.
}
\label{fig:4body1}
\end{figure}

The fourth order increments are of course much more numerous. Adopting the same cutoff criterium as in the case of the two-body terms, one would obtain $71$ structures to be considered. However, taking into account our conclusion regarding the convergence of the three-body increments, we have restricted the examination of four-body increments to the cases containing $r_{NN}=a/\sqrt{2}$. Among them are:  (i) \textit{connected} structures, where $r_{12}=r_{23}=r_{34}=a/\sqrt{2}$ [e.g. Figure \ref{fig:4body1} (a-i)], $19$ in total, referred to as [4] further in the text; (ii) structures, where three atoms are connected and the distance between the fourth atom and one of the three connected atoms is not longer than the lattice constant $a$  [e.g. Figure \ref{fig:4body1} (j,k)], $12$ in total, referred to as [3]+1 further in the text; (iii) structures, where $4$ atoms are connected by pairs and distance between pairs is not longer than the lattice constant $a$  [e.g. Figure \ref{fig:4body1} (l-n)], $6$ in total, referred to as [2]+[2] further in the text. As can be expected, $\sum\Delta\varepsilon_{ijkl}^{[4]}>\sum\Delta\varepsilon_{ijkl}^{[3]+1}>\sum\Delta\varepsilon_{ijkl}^{[2]+[2]}$ (Tables \ref{ca-corr} and \ref{sr-corr}). The same is true for the individual contributions belonging to these groups. The absolute values of the individual increments are smaller than the related three-body terms independently on the group and the method employed. 

\begin{figure}
\includegraphics[width=1.00\textwidth]{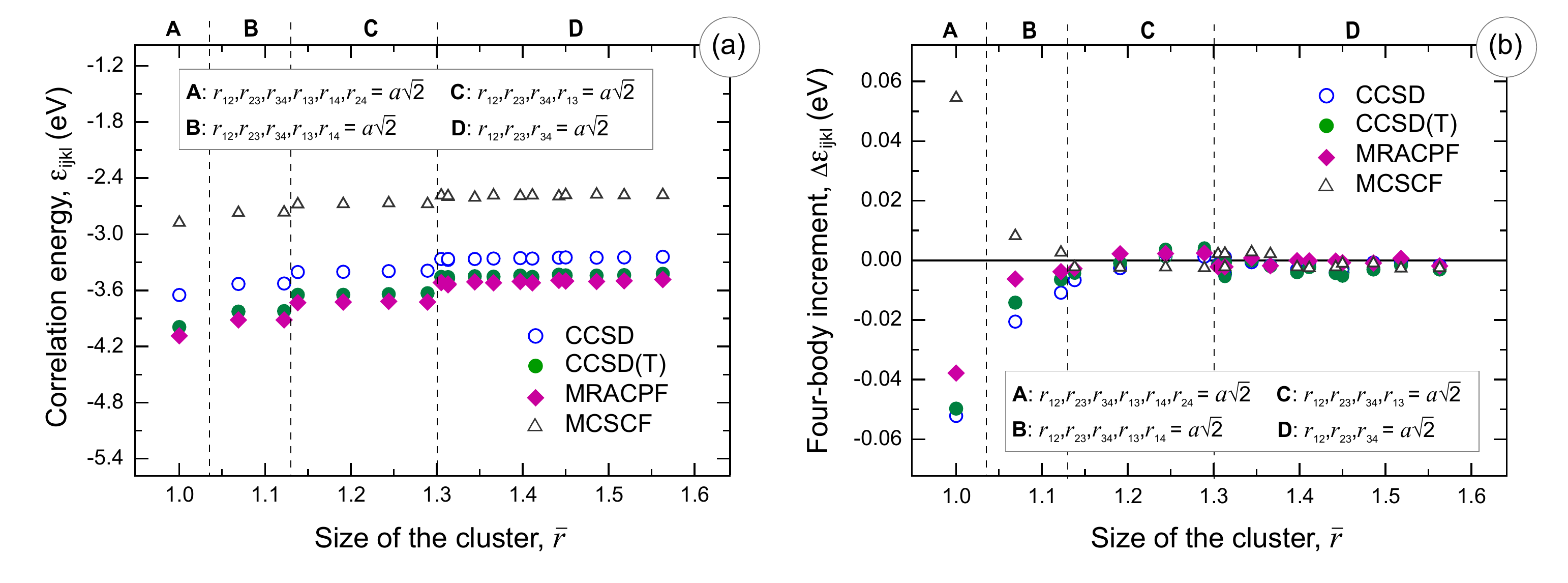}
\caption{Correlation energy contributions ($\varepsilon_{ijkl}^{[4]}$ and $\Delta\varepsilon_{ijkl}^{[4]}$) are plotted as functions of the size of the \textit{connected} four-body clusters as obtained for Ca at different levels of theory (see text for details).  A measure for the size of the cluster is an arithmetic mean of the distances between metal atoms in the central part, $\bar{r}=\sum r_{ij} /6$, in units of $a/\sqrt{2}$. The respective pictures obtained for Sr are qualitatively the same.
}
\label{fig:4body2}
\end{figure}

In Figure \ref{fig:4body2} correlation energies, $\varepsilon_{ijkl}^{[4]}$, as well as the corresponding non-additive four-body contributions, $\Delta\varepsilon_{ijkl}^{[4]}$, are plotted as functions of the size of the considered clusters.  As in the case of the three-body contributions,  the weight of the static correlations increases with decreasing number of nearest-neighbor distances ($r_{NN}=a/\sqrt{2}$) within one structure. Since in the case of the four order increments we are focusing only on compact structures,  
the MCSCF energy covers $70$-$76$\,\% of the MRACPF result as it was observed for the compact three-body cases. Comparing Figures \ref{fig:3body2} and \ref{fig:4body2}, one can identify further similar features of the three- and four-body contributions.  Thus, $\Delta\varepsilon_{ijkl}$ are higher in magnitude for the more compact structures. Consistently with the fact that the [3]+1 and [2]+[2] structures contain by definition not more than $3$ nearest-neighbor distances, the related contributions are small and the sums are non-negligible only due to the high weight factors associated with the most compact members of each group. It is interesting to note, that while $\sum\Delta\varepsilon_{ijkl}^{[4]}<0$, the related contributions from the [3]+1 and [2]+[2] are repulsive. Though consisting approximately one-third of the corresponding four-body sum, these two groups cannot compensate negative contribution $\sum\Delta\varepsilon_{ijkl}^{[4]}$ obtained with the CC methods. The resulting four-body sum is thus too high in magnitude and $\sum\Delta\varepsilon_{ijkl} > \sum\Delta\varepsilon_{ijk}$ or $\sum\Delta\varepsilon_{ijkl} \approx \sum\Delta\varepsilon_{ijk}$ in the case of CCSD(T) or CCSD, respectively. At the MRACPF level this convergence criterium is, however, satisfied and $\sum\Delta\varepsilon_{ijkl}$ consists less than $10$\,\% of $\sum\Delta\varepsilon_{ijk}$. 

\begin{figure}[t]
\includegraphics[width=1.0\textwidth]{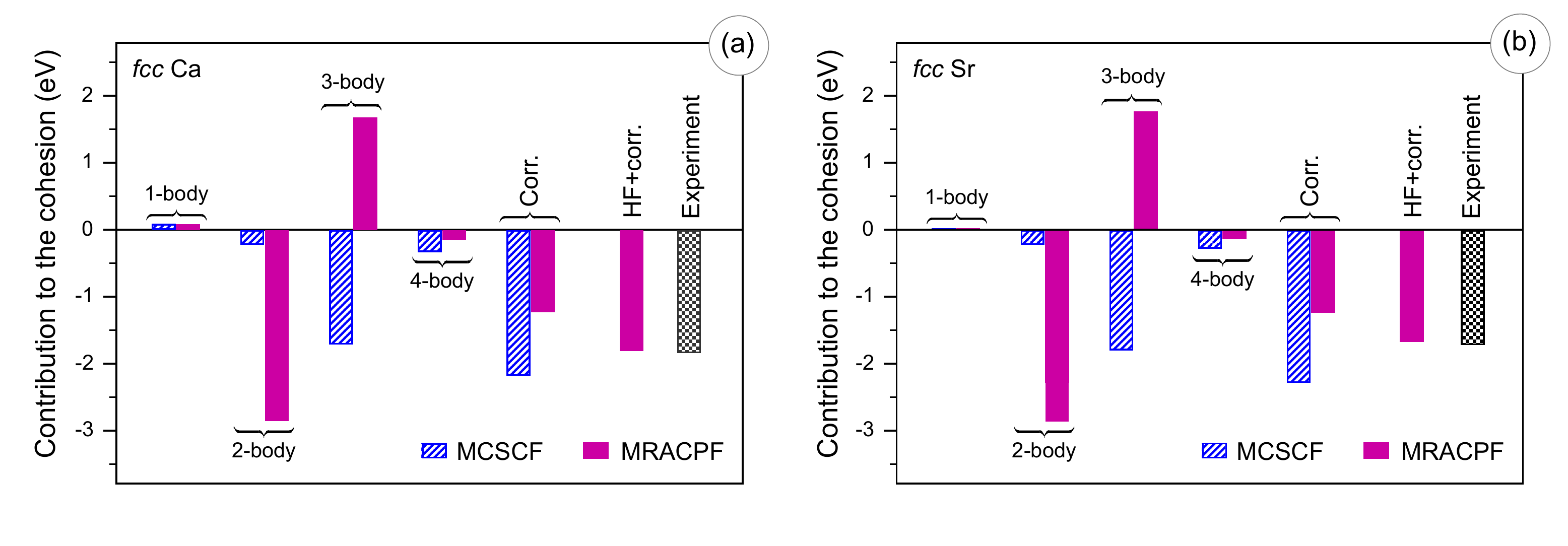}
\caption{Cohesive energies of the \textit{fcc} Ca (a) and \textit{fcc} Sr (b): one-, two-, three-, and four-body contributions obtained at different levels of theory and the calculated cohesive energy as compared to the experimental value.}
\label{fig:diag}
\end{figure}

Taking into account the failure of convergence in the case of the SR methods tested in this work, for further analysis as well as comparison of the calculated and experimental cohesive energy we focus only at the data obtained with the MR approach. In the discussion above we have shown how the static and dynamic correlations contribute to the individual increments of the same order. Now we can compare their contributions to the total sums (see Tables \ref{ca-corr} and \ref{sr-corr} as well as Figure \ref{fig:diag}). Thus, the importance of static correlations grows with the order of the increments. $\sum\Delta\varepsilon_{ij}$ is determined by dynamic correlations and multi-configurational constituent is very small in this case. Contrary, the MCSCF contribution to the three-body term is large and attractive that is due to an increase of the metallic character when turning from the two-body clusters to the three-body ones. The contribution of dynamic correlations to the three-body term evaluated at the MRACPF level is repulsive, allowing to compensate the overestimation at the two-body level. The latter is due to the too local character of the wave function evaluated at the two-body level as compared to the one of the real metal. Overall, at this stage, i.e. after accounting for the third order correction, one gets a balanced description of metallicity and electron correlation in the studied systems. Consequently, the four-body term is small due to the cancellation of (small) repulsive and attractive non-additive contributions. 

It is interesting to note, that due to the growing metallic character when going from the studied earlier group 12 metals~\cite{zn,hg} and Mg~\cite{mg1,mg2} to Ca and Sr, the role of \textit{non-compact} increments is increased. Thus, in the former case it was sufficient to consider only compact three-body structures which are shown in Figure \ref{fig:3body1} (a-h), whereas in the present work the amount of the three-body clusters considered is significantly increased in order to achieve the sufficient level of delocalisation. Furthermore, comparing Ca and Sr, on one hand, and Mg, on the other hand, one can realise significant difference even at the level of the two-body increments. Whereas the difference between two-body contribution to the correlation energy computed with and without perturbative triples in the case of \textit{hcp} Mg consists ca.\,\,$10$\,\%, for the two metals studied in this work inclusion of perturbative triples yields increase of the correlation energy by almost $30$\,\%. Such result already indicates rather strong multi-reference character in the latter case and suggests that CCSD(T) may diverge as, in fact, was observed in the present work. This observation is in good agreement with the results obtained by Gr\"uneis and co-workers when applying wave-function-based correlation methods to the electron gas~\cite{gruen}.

Overall, summing up all correlation contributions obtained with MRACPF and addition of the corresponding HF values allows to achieve very good agreement between computed and experimental cohesive energies for both Ca and Sr (Figure \ref{fig:diag}), covering approximately $97$\,\% of the experimental correlation energy for the both studied systems. Our estimate for errors due to truncation of the incremental expansion is about $\pm 3$\,\%. A basis set extrapolation at the correlated level would yield an increase of the cohesive energy by about $1$-$2$\,\%. Also the basis set limit is probably not reached at the HF level, so we can expect a change of the HF cohesive energy by $1$-$2$\,\% as well.  
  
\section{Conclusions}

We have applied an incremental scheme to calculate correlation energy of the bulk metals - calcium and strontium. In the case of metals, the incremental expansion is based on a well-localizable model system rather than on the real metal and to allow for delocalization (and thereby approaching the real metal) only gradually within the various levels of the many-body expansion. This approach makes sense only if the number of non-negligible terms is sufficiently small. At the coupled-cluster level the usual criterion for the convergence of the incremental scheme is not fulfilled as the sum of three-body increments is, at best, comparable in magnitude with the corresponding fourth order correction. Thus, both the CCSD method and the augmented CCSD(T) theory, forming a wave function through excitations from a single reference Slater determinant, fail to describe calcium and strontium as the related wave functions include  several determinants contributing with comparable weights.  Consequently, the situation is much better when dealing with an approximately size-extensive multi-reference correlation method, the averaged coupled pair functional. Overall, one gets a balanced description of metallicity in the studied systems at the levels of three-body increments. Although three-body increments are as high as $60$\,\% of the sum of two-body terms, four-body increments are $8$ times smaller. Summing up all correlation contributions obtained with MRACPF, one gets almost full amount of the experimental correlation energy. 

\section*{Acknowledgements}
The authors would like to thank Professor H. Stoll (Stuttgart) and Professor K. Rosciszewski (Krakow) for useful discussions. 
EV appreciates the support from the German Research Foundation (DFG) through the project VO 1711/2-1 ``Local correlation method for metals''. Worthwhile suggestions from PD Dr.~Yu.~S.~Dedkov (Berlin) during preparation of the manuscript are acknowledged.

\end{document}